\documentclass[final,1p,times]{elsarticle} 
\usepackage{graphicx} 
\usepackage{amssymb} 
\usepackage{amsthm} 
\usepackage{lineno} 

\journal{Nuclear Physics A} 
\begin{document} 

\begin{frontmatter} 


\title{\bf {Addressing the physics of the ridge by 2- and 3-particle correlations at STAR}}

\author{\rm Pawan Kumar Netrakanti for the STAR Collaboration}

\address{Physics department, Purdue University, West Lafayette-47906, IN USA}

\begin{abstract} 
We present new results on 2-particle azimuthal ($\Delta\phi$) correlation relative 
to event plane and 3-particle pseudorapidity ($\Delta\eta$) correlation at mid-rapidity
in Au+Au collisions at $\sqrt{{\it s}_{NN}}$ = 200 GeV, measured by the STAR experiment. 
While jet-like correlation is symmetric, ridge is 
found to be asymmetric when trigger particle azimuth 
is between in- and out-of-plane.
The charge ordering properties between associated and trigger particles
are exploited to separate jet-like and ridge contributions in 
3-particle $\Delta\eta$-$\Delta\eta$ correlations. We found that like-sign
triplets are dominated by ridge. The separated ridge, while narrow 
in $\Delta\phi$, is extremely broad in $\Delta\eta$. The ridge particles
are not only uncorrelated to the trigger particle in $\Delta\eta$, but
also uncorrelated between themselves.
\end{abstract} 

\end{frontmatter} 






\section{Introduction}
The observation of the ridge, a long range
pseudorapidity ($\Delta\eta$) correlation at near-side~\cite{RIDGE}, 
in dihadron correlations has motivated many 
theoretical investigations~\cite{theory,Hwa}. Understanding 
the formation mechanism of the ridge may provide insight 
into the early dynamics of relativistic heavy ion collision.
We investigate the ridge using two tools: dihadron correlation relative to
event plane and 3-particle $\Delta\eta$-$\Delta\eta$ correlations.
We discuss what have already been learned from previous studies~\cite{Aoqi,QM08} 
and present new findings at this conference.

The ridge magnitude was found to decrease with trigger particle azimuth
relative to event plane ($\phi_{s}=\phi_{trig}-\psi_{EP}$) 
from in-plane ($\mid$$\phi_{s}$$\mid$$\sim$$0^{\circ}$) to 
out-of-plane ($\mid$$\phi_{s}$$\mid$$\sim$$90^{\circ}$)~\cite{Aoqi}. 
Motivated by our data, Chiu and Hwa~\cite{Hwa} 
proposed the Correlated Emission Model (CEM) where the ridge is 
formed by correlated particle emission
due to aligned jet propagation and medium flow and predicted
asymmetric near-side $\Delta\phi$ correlations. 
We present a study of this asymmetry from our data.
We found the ridge is asymmetric and the asymmetry 
peaks at $\phi_{s}$$\sim$$45^{\circ}$.

The broadness of the ridge was found to be event-by-event 
from 3-particle $\Delta\eta$-$\Delta\eta$
correlations~\cite{QM08}. By further analyzing their charge dependence, 
we may separate jet-like and ridge components without assuming 
the shape of the ridge in $\Delta\eta$.
We find like-sign triplets are dominated by ridge and use this observation
to separate the jet-like and ridge components. We find jet-like correlation is narrow, and ridge
is broad and approximately uniform in $\Delta\eta$.

\section {Dihadron correlation relative to event plane}
Dihadron correlations are studied for trigger particles 
separately in $\phi_{s}$$>$$0^{\circ}$ and $\phi_{s}$$<$$0^{\circ}$ for 
20-60$\%$ Au+Au collisions at $\sqrt{{\it s}_{NN}}$ = 200 GeV. 
The $p_{T}$ ranges of the trigger and associated particles are 
3$<$$p_{T}^{trig}$$<$4 GeV/c and 1$<$$p_{T}^{assoc}$$<$ 2 GeV/c, 
respectively, and both within $\mid$$\eta$$\mid$$<$1.
The event plane angle is constructed from particles outside the 
$p_{T}$ range used in the correlation study.
The dihadron correlation signal obtained for each $\phi_{s}$ slice
has a large background contribution from anisotropic 
flow ($v_{2}$ and $v_{4}$).
The construction and subtraction of this background is described in detail
in Ref.~\cite{NIM}.
The jet-like yield is obtained from the difference 
between the raw correlation in $\mid$$\Delta\eta$$\mid<$0.7, 
containing both jet-like component
and the ridge, and that in $\mid$$\Delta\eta$$\mid>$0.7 
(scaled by $\Delta\eta$ acceptance factor) containing
presumably only the ridge component.
The systematic uncertainties are largely
canceled in the jet-like result.
To obtain the near-side ridge, we fit the away-side
correlation within $\mid$$\Delta\eta$$\mid>$0.7 to two Gaussian of
the same width, and then subtract the extrapolated fit to the near-side
to remove away-side leakage to the ridge~\cite{Josh}.
The remaining is the near-side ridge $\Delta\phi$ correlation.

We analyze the asymmetry parameter of the $\Delta\phi$ correlations by 
$A = \frac{N_{(0,1)} - N_{(-1,0)}}{N_{(0,1)} + N_{(-1,0)}}$ where $N_{(a,b)}=\int_{a}^{b}\frac{1}{N_{trig}}\frac{dN}{d\Delta\phi}d\Delta\phi$.  
\begin{figure}[htp]
\vspace{-0.1in}
\begin{minipage}{0.5\textwidth}
\includegraphics[height=10pc,width=14pc]{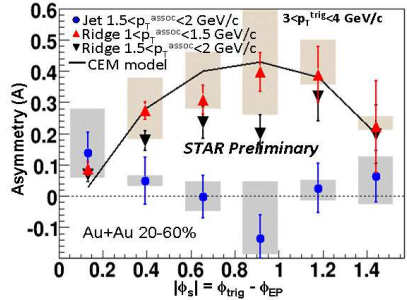}
\end{minipage}
\begin{minipage}{0.5\textwidth}
\caption{
Asymmetry parameter for jet-like and ridge azimuthal correlations 
as a function of $\phi_{s}$ for 
20-60$\%$ Au+Au collisions. The shaded bands represent the 
systematic uncertainties in the flow estimation. Also shown  
by the black curve is the prediction from CEM.
}
\label{Fig1}
\end{minipage}
\end{figure}
Figure~\ref{Fig1} shows $A$ for jet-like and ridge 
$\Delta\phi$ correlations as a function of $\phi_{s}$ for 20-60$\%$ 
Au+Au collisions.
The ridge is found to be asymmetric with the maximum asymmetry 
appearing at $\phi_{s}$$\sim$45$^{\circ}$, while jet is symmetric. 
The ridge asymmetry is 
persistent for different associated $p_{T}$. 
The ridge asymmetry is qualitatively consistent with the 
CEM prediction~\cite{Hwa}, suggesting that formation of the ridge may be 
due to jet-flow alignment.
However, we note that CEM does not have dynamics depending on particle
$p_{T}$, nor does it have mechanisms to create correlations 
between particles separated by large $\Delta\eta$.
In general, the ridge $\Delta\phi$ asymmetry may be obtained 
by models which incorporate interplay between jet propagation and
radial flow.

\section{Three particle $\Delta\eta$-$\Delta\eta$ correlations}
We have analyzed the 3-particle $\Delta\eta$-$\Delta\eta$ correlations data 
in $d$+Au, 40-80$\%$ and 0-12$\%$ Au+Au collisions 
at $\sqrt{{\it s}_{NN}}$ = 200 GeV~\cite{QM08}. The trigger and associated 
particles are restricted to $\mid$$\eta$$\mid$$<$1 and 
$\mid$$\Delta\phi$$\mid$$<$0.7. 
The $p_{T}$ ranges are 1$<$$p_{T}^{assoc}$$<$3$<$$p_{T}^{trig}$$<$10 GeV/c. 
We further analyze different charge combinations in an attempt 
to separate the jet-like and ridge components and their structures. 
The 3-particle correlation raw signal is obtained from all triplet of 
one trigger and two associated particles from the same 
event. The signal is binned in $\Delta\eta_{1}$ and 
$\Delta\eta_{2}$, the pseudorapidity differences between the 
associated particles and the trigger. Combinatorial background arises 
when one or neither of the two associated particles are correlated 
with the trigger particle besides flow correlation. Details about background 
construction, systematic uncertainties and correction factors 
can be found in Ref.~\cite{QM08}.
The 2- and 3-particle correlation yields are corrected 
for the centrality-, $p_{T}$- and $\phi$-dependent reconstruction efficiencies for 
associated particles and the $\phi$-dependent efficiency for trigger particles,
and are normalized per corrected trigger particle.
\begin{figure}[htp]
\vspace{-0.1in}
\begin{minipage}{0.5\textwidth}
\includegraphics[height=10pc,width=14pc]{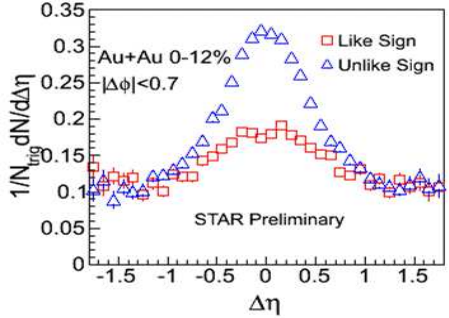}
\end{minipage}
\begin{minipage}{0.5\textwidth}
\caption{
Background subtracted dihadron correlation in 
$\Delta\eta$ ($\mid$$\Delta\phi$$\mid$$<$0.7)
corrected for $\Delta\eta$ acceptance in 0-12$\%$
Au+Au collisions for 3$<$$p_{T}^{trig}$$<$10 GeV/c and 
1$<$$p_{T}^{assoc}$$<$3 GeV/c. The distributions are separated between 
like-sign (square) and unlike-sign (triangle) trigger-associated pairs.
}
\label{Fig2}
\end{minipage}
\end{figure}

Figure~\ref{Fig2} shows the dihadron $\Delta\eta$ correlation between
associated and trigger particles on the near-side 
($\mid$$\Delta\phi$$\mid$$<$0.7) in 0-12$\%$ Au+Au collisions.
The correlation is separated for like-sign and unlike-sign pairs. 
There are more unlike-sign than like-sign 
pairs at $\mid$$\Delta\eta$$\mid$$\sim$0. This is due to the charge ordering 
in jet fragmentation. At larger $\mid$$\Delta\eta$$\mid$$>$0.7, the
unlike-sign and like-sign correlations are similar indicating the 
ridge particles have no charge dependence. 
Correlation in same-sign triplets should be dominated by ridge correlation
because the probability of a jet fragmenting into 3 same-sign 
particles in our $p_{T}$ range is negligible. Therefore, we separate
the charge ordered triplets in 3-particle $\Delta\eta$-$\Delta\eta$ 
correlations with like-sign triplets ($AA^{like}T^{like}$) 
and like-sign associated pairs with 
an opposite-sign trigger particle ($AA^{like}T^{unlike}$).

\begin{figure}[htp]
\vspace{-0.1in}
\centering
\includegraphics[height=12pc,width=28pc]{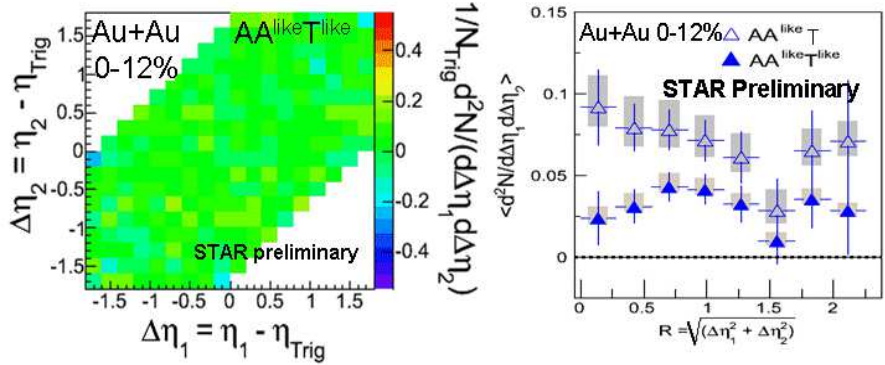}
\caption{
Left: Background subtracted 3-particle $\Delta\eta$-$\Delta\eta$ 
correlations for 3$<$$p_{T}^{trig}$$<$10 GeV/c and 
1$<$$p_{T}^{assoc}$$<$3 GeV/c 
for $AA^{like}T^{like}$ in 0-12$\%$ Au+Au collisions. 
Right: Average 3-particle correlations signal
for associated particles are of same charge sign as a function of
$R$ on $\Delta\eta$-$\Delta\eta$ plane in 0-12$\%$ Au+Au collisions.
Systematic errors are shown in the 
shaded box due to background normalization 
}
\label{Fig3}
\end{figure}

Figure~\ref{Fig3}(left) shows the background subtracted 
3-particle $\Delta\eta$-$\Delta\eta$ correlations for $AA^{like}T^{like}$
in 0-12$\%$ Au+Au collisions. 
The average correlation signal seems to be uniform over 
$\Delta\eta$-$\Delta\eta$ region. To quantify this we 
study the average signal as a function of $R$ on $\Delta\eta$-$\Delta\eta$ plane, shown in 
Fig~\ref{Fig3}(right) for 
$AA^{like}T$ ($AA^{like}T^{like}$+$AA^{like}T^{unlike}$) and 
$AA^{like}T^{like}$ in 0-12$\%$ Au+Au collisions.
$AA^{like}T$ contains both jet-like and ridge contributions while $AA^{like}T^{like}$ 
is dominated by ridge.
Since the ridge at large $\Delta\eta$ is identical for like-sign and unlike-sign trigger-associated 
pairs in dihadron correlations, it should also be the same at large $\Delta\eta$ in $AA^{like}T^{like}$ 
and $AA^{like}T^{unlike}$ because the change from the dihadron to 
trihadron correlation is the addition of third particle of the same charge sign.
\begin{figure}[htp]
\vspace{-0.1in}
\begin{minipage}{0.5\textwidth}
\includegraphics[height=10pc,width=14pc]{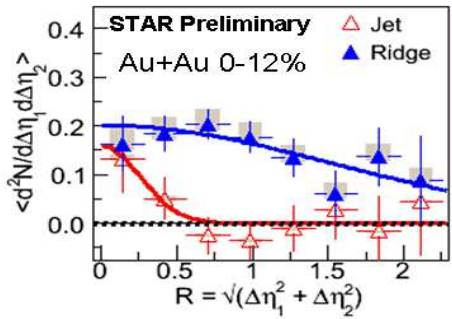}
\end{minipage}
\begin{minipage}{0.5\textwidth}
\caption{
Average signal of the jet-like and ridge as a function of $R$ 
on $\Delta\eta$-$\Delta\eta$ plane for 3$<$$p_{T}^{trig}$$<$10 GeV/c and 
1$<$$p_{T}^{assoc}$$<$3 GeV/c in 0-12$\%$ Au+Au collisions. 
The curves are Gaussian fits to the data.
}
\label{Fig4}
\end{minipage}
\end{figure}
Therefore, the total jet-like component can be obtained by:
Jet = 3($AA^{like}T^{unlike}$ - $AA^{like}T^{like}$),
where the factor of 3 takes into account the jet contribution from 
$AA^{like}T^{unlike}$ and $AA^{unlike}T$ charge combinations.
The ridge can then be obtained by subtracting the jet-like
component from the total 3-particle correlation where all 
charge combinations are included. Alternatively the ridge
can be taken simply as 4 times $AA^{like}T^{like}$. The
two methods yield consistent results.

Figure~\ref{Fig4} shows the average signal of the separated ridge and 
jet-like components from 3-particle $\Delta\eta$-$\Delta\eta$ correlations 
as a function of $R$ on $\Delta\eta$-$\Delta\eta$ plane for 0-12$\%$ Au+Au collisions.
The jet-like component is narrow and symmetric 
between $\phi$ and $\eta$ having a Gaussian width 
$\sigma_{\Delta\eta}\sim$0.25$\pm$0.09. 
The separated ridge, while narrow in $\Delta\phi$, is 
extremely broad with Gaussian width 
$\sigma_{\Delta\eta}\sim$1.53$\pm$0.41.
The ridge is also consistent with a uniform distribution 
within our acceptance with reasonable $\chi^{2}$/NDF.
\section{Summary}
In summary, we have presented dihadron correlations relative to 
event plane 
for Au+Au 20-60$\%$ collisions. 
While jet-like correlation is symmetric, ridge is 
found to be asymmetric when trigger particle azimuth 
is between in- and out-of-plane. The asymmetry is 
qualitatively consistent with the correlated particle 
emission model prediction, suggesting the formation mechanism of
the ridge may be due to alignment of jet propagation and medium flow.

We have used the charge ordering properties to separate the 
jet-like and ridge components via 3-particle $\Delta\eta$-$\Delta\eta$ correlations 
in 0-12$\%$ Au+Au collisions. This stems out of the observation that 
like-sign triplets are dominated by the ridge. 
We found that the jet-like correlation is narrowly confined, and the ridge is broadly distributed 
being approximately uniform in $\Delta\eta$ in our measured acceptance. 
Except small angle correlation in azimuth, the ridge particles 
appear to be uncorrelated not only with the trigger particle, 
but also between themselves in $\Delta\eta$. 

\end{document}